\documentclass[12pt,preprint]{aastex}

\shorttitle{A GPU-Enabled Microlensing Parameter Survey}
\shortauthors{Bate et al.}

\begin{document}


\title{A GPU-Enabled, High-Resolution Cosmological Microlensing Parameter Survey\altaffilmark{*}}


\author{N. F. Bate$^{1,2}$ and C. J. Fluke$^1$}
\affil{$^1$Centre for Astrophysics \& Supercomputing, Swinburne University
of Technology, PO Box 218, Hawthorn, Victoria, 3122, Australia}
\affil{$^2$Sydney Institute for Astronomy, School of Physics, A28, University of Sydney, NSW, 2006, Australia}
\email{cfluke@swin.edu.au}

\altaffiltext{*}{Research undertaken as part of the Commonwealth
  Cosmology Initiative (CCI: www.thecci.org), an international
  collaboration supported by the Australian Research Council.}

\begin{abstract}
In the era of synoptic surveys, the number of known gravitationally
lensed quasars is set to increase by over an order of magnitude.  
These new discoveries will enable a move from single-quasar studies 
to investigations of statistical samples, presenting new 
opportunities to test theoretical models for the structure of quasar accretion discs and broad emission line regions (BELRs).  As one crucial step in preparing
for this influx of new lensed systems, a large-scale exploration 
of microlensing convergence-shear parameter space is warranted, requiring
the computation of {\bf$O(10^5)$} high resolution magnification maps. 
Based on properties of known lensed quasars,
and expectations from accretion disc/BELR modelling, we identify 
regions of convergence-shear parameter space, map sizes, smooth matter fractions, and 
pixel resolutions that should be covered.  We describe how the 
computationally time-consuming task of producing {\bf$\sim290000$} magnification 
maps with sufficient resolution ($10000^2$ pixel/map) to probe scales
from the inner edge of the accretion disc to the BELR can 
be achieved in $\sim400$ days on a 100 teraflop/s high performance computing
facility, where the processing performance is achieved with graphics processing
units. We illustrate a use-case for the parameter survey by investigating
the effects of varying the lens macro-model on accretion disc constraints
in the lensed quasar Q2237+0305. We find that although all constraints are consistent within their current error bars, models with more densely packed microlenses tend to predict shallower accretion disc radial temperature profiles. With a large parameter survey such as the one described here, such systematics on microlensing measurements could be fully explored.
\end{abstract}


\section{Introduction}
Cosmological gravitational microlensing refers to the regime of gravitational
lensing where individual stars within a lensing galaxy act to magnify 
a background source, usually a quasar.  The most obvious effect 
of microlensing is to produce an uncorrelated brightness change with time
in a single image of a multiply-imaged source; an intrinsic variation in 
source flux would appear in all images separated by the individual 
image time delays.
Cosmological microlensing was first observed in the lensed quasar
\object{Q2237+0305} by \citet{irwin+89}, and it is likely to be taking place on some level in all quasars 
multiply imaged by a foreground galaxy. For
recent summaries, see \citet{wambsganss06} and \citet{schmidt+10}.

While gravitational microlensing is achromatic, the magnitude of
microlensing-induced fluctuations depends strongly on
the size of the emission region. Small sources (relative to the
Einstein radius of the microlenses) are affected more strongly than larger
sources. Detections of microlensing signals in a single waveband can
therefore be used to constrain the size of the quasar emission
region. For example, optical observations of \object{Q2237+0305} suggest a
V-band accretion disc radius of $R_V =
6.2^{+3.8}_{-2.7}\times10^{15}$cm \citep{pk10}.

Multi-wavelength observations present an opportunity to constrain 
the wavelength-dependent structure of quasar emission
regions. First constraints on the temperature profile of quasar
accretion discs have already been published (\citealt{anguita+08};
\citealt{bfww08}; \citealt{eigenbrod+08b}; \citealt{pmk08};
\citealt{fbw09}; \citealt{blackburne+11}; \citealt{mediavilla+11}; \citealt{mosquera+11a}). 

Interpretation of broad emission line region (BELR) signals
are more complex, however some early attempts have been made. \citet{lewis+98} and \citet{wow05} measured the relative sizes of the CIII]/MgII BELR and continuum emission region in Q2237+0305 using microlensing observations. \citet{abajas+07} proposed a biconical BELR as a possible explanation for a recurrent blue-wing enhancement in the emission lines of SDSS J1004+4112 (\citealt{richards+04};
\citealt{gomez+06}). \citet{odowd+10} compared a differential microlensing signal observed across the 
CIII] emission line in Q2237+0305 with various BELR geometries, and \citet{sluse+11} studied the broad emission lines in 39 epochs of Q2237+0305 data obtained with the VLT.

While  $\sim100$ multiply imaged quasars have now been 
discovered,\footnote{See the CASTLES Survey webpage:
{\tt http://www.cfa.harvard.edu/glensdata/}} only $\sim20$ of these have
measured light travel time delays between lensed images -- 
see \citet{oguri07} for a recent compilation. 
New, high-cadence, all-sky surveys such as those planned for the SkyMapper Telescope \citep{keller+07} and the Large 
Synoptic Survey Telescope
(LSST; \citealt{ivezic+08}) are likely to discover a large number
of gravitationally lensed quasars. For example, \citet{oguri+10}
predict that LSST will find approximately 8000 new lensed quasars
across its 10-year 20000-deg$^2$ survey, 3000 of which will have 
good measurements of the time delays between images.

The anticipated increase in the number of known lensed quasars
will take measurements of quasar structure beyond the realm of 
single-object studies and into that of statistically
meaningful samples. For example, we can test whether a single accretion mechanism is consistent with all observations, examine scaling relations between black hole mass and accretion disc radii [see \citet{morgan+10} for such an analysis with a sample of 11 lensed quasars], and investigate secondary parameters such as Eddington ratio \citep[e.g][]{mosquera+11b}.

Analysis and interpretation of this wealth of new information regarding 
quasar emission regions will require extensive simulations.  The basic tool 
for microlensing is the magnification map: a pixellated 
approximation to a region of the (background) source plane, with 
a magnification value defined at each physical source location. By 
convolving magnification maps with wavelength-dependent geometric model 
profiles, both simulated lightcurves and 
statistical magnification distributions can be extracted, 
ready for comparison with observational data.

The characteristic caustic structure of a magnification map depends on
the convergence and shear (described in Section \ref{sec:survey}). 
These parameters are local properties of the global macro-lens model 
at the position of each lensed image. Since each lensing environment 
is unique, potentially $\sim 6800$ new combinations of convergence and 
shear will need to be simulated [\citet{oguri+10} predict $\sim14$ per cent of 
$\sim 3000$ LSST-discovered systems will be quadruply imaged, with the remaining 
systems dominated by double images, resulting in a mean image number of 2.28].
Moreover, macro-lens models are not uniquely determined. For example, \citet{wms95} 
modeled lensing systems with three different lens potentials and \citet{waf02} 
contains a compilation of lens models for \objectname{Q2237+0305} by eight 
different authors: each of which predicts different convergence and shear values 
for the lensed images.   The potential need for multiple models, and hence
mangification maps, for each newly discovered system further increases the 
expected coverage of convergence-shear parameter space. 

Although the computational time needed for analysis of magnification maps 
can be considerable \citep[e.g.][]{pk10a}, generating sufficient high-resolution, statistically 
useful magnification maps is itself a computationally-demanding task.  
Over the last two decades, the most widely-used approaches 
to generating magnification maps are based on backwards ray-shooting 
(described in more detail in Section 2). Here, light rays are propagated
from the observer, through  the lens plane, requiring the calculation of
deflections by $N_*$ individual lenses, and mapped onto the pixellated
source plane. This computation is repeated for a sufficiently 
large number of light rays, typically $O(10^6)$ or higher, in order to
achieve statistically significant coverage per pixel in the source plane.  

In the hierarchical tree method (\citealt{wambsganss90};
\citealt{wambsganss99}), individual lenses are approximated 
by higher-mass pseudo-lenses, depending on their distance from a given 
light ray, thus avoiding the requirement for $N_*$ deflection
calculations per light ray.  \citet{k04} describes a computationally-efficient approach
to reducing the $N_*$ effect, where the lens and source planes are treated as spatially periodic,
allowing the use of Fourier methods (similar to the P$^3$M methods
used for many-particle gravitational force computations). \citet{Mediavilla06} and
\citet{Mediavilla11} adaptively tessellate the lens plane, with the
lens equation used to map the lattice of polygons from the image to source plane,  
where the polygons are efficiently mapped onto source pixels. This approach reduces 
the total number of light rays required to reach a given statistical accuracy 
by over three orders of magnitude. 

The emergence of the graphics processing unit (GPU) as a means of 
accessing low cost, high performance, massively parallel computing, 
presents an alternative opportunity for speeding-up deflection angle 
calculations with an algorithmically-simple solution. Astronomers have been amongst the early adopters of GPUs for scientific computation, 
reporting typical speed-ups between $\times 30$ to $\times 100$ over what they 
can achieve on single (or few) core CPUs.  While further work is required to fully understand
which types of computations are best suited for GPUs [see \citet{barsdell+10}
and \citet{fluke+11} for a discussion of these issues specifically
aimed at astronomers], algorithms that exhibit the properties of high
arithmetic intensity and a high-level of data parallelism are the most
promising targets.

\citet{thompson+10} demonstrated that by taking advantage of the highly 
parallel GPU-architecture, a ``brute force'' implementation of ray-shooting 
is now a practical alternative.  Moreover, Bate et al. (2010) have shown 
that on current generation multi-GPU systems (e.g. the 4-GPU NVIDIA Tesla S1070 
unit with a processing peak of $\sim 2.5$ teraflop/s for
single precision computation\footnote{1 teraflop/s = $10^{12}$ floating
point operations per second.}), the direct code on GPU executes in runtimes
comparable too, and often faster than, the single-core tree code.  

Our target system is the Australian GPU Supercomputer for Theoretical Astrophyiscs 
Research (gSTAR\footnote{The gSTAR facility was part-funded by Astronomy Australia Limited 
through 
the Australian Federal Government's Education Investment Fund.}), a hybrid
CPU+GPU cluster containing 102 $\times$ NVIDIA C2070 GPUs 
and 21 $\times$ M2090 GPUs, connected via QDR Infiniband, with a (theoretical,  
single-precision) peak of $\sim 130$ teraflop/s. 
Throughout this work, we estimate run-times for two different
GPU cluster configurations: we assume a slightly more conservative
100 teraflop/s mode of operation for the phase one gSTAR system, gSTAR$_{100}$, 
and a somewhat optimistic future upgrade with a peak of 600 teraflop/s, gSTAR$_{600}$,
based on the original system design goals. 
In both cases, we expect to achieve at least 50 per cent of the theoretical peak, 
based on the benchmarking in \citet{thompson+10}.

Our goal in this paper is to demonstrate that a high resolution,
large-scale, microlensing parameter survey is feasible on a GPU-enabled computing 
cluster.  This requires generating statistically useful samples of high-resolution 
magnification maps over the entire range of convergence-shear parameter space -- a theoretical
equivalent of an ``all-sky'' survey.  By providing {\bf$\sim 290000$} pre-computed magnification maps as a Web-accessible resource, rapid first look computations would be possible as new  
microlensed quasars are discovered.  High-resolution coverage of parameter space 
enables investigation of degeneracies between macro-model parameters and inferred 
plausible models for quasar emission regions, and presents new opportunities to test 
(e.g. via mock observations) theoretical models of the accretion disc and BELR 
well before the new era of microlensed quasar discoveries begins.
A parameter survey does not remove the need for detailed modelling of individual systems:
for example, simultaneous multi-image fits to light curves including dynamical motions of 
the stellar lens populations, requiring generation of magnification maps for each observation
epoch, as per \citet{pk10a}.

This paper is organised as follows. In Section \ref{sec:survey}, we discuss the theoretical
considerations  for a survey of the entirety of convergence-shear
parameter space on a GPU-based computing cluster. We discuss magnification map size and resolution
considerations, and use simulation time estimates based on the results of
\citet{thompson+10} to inform a realistic simulation strategy.  In Section \ref{sec:first_app} 
we illustrate one possible use for our survey by examining the systematic
effects on accretion disc constraints in the lensed quasar
\objectname{Q2237+0305} when the lens macro-model is varied. We present our
conclusions in Section \ref{sec:conc}.

Throughout this paper, a cosmology with $H_0 = 70\rm{kms^{-1}}\rm{Mpc^{-1}}$, $\Omega_m = 0.3$ and $\Omega_\Lambda = 0.7$ is assumed.

\section{Survey design}
\label{sec:survey}

In this section, we discuss the theoretical considerations that inform the choices of parameters 
for a high-resolution, cosmological microlensing parameter space survey. We discuss the ways 
in which observational constraints shape the magnification map sizes and resolutions that are 
required to produce a useful resource for the microlensing community.

There are two techniques for constraining quasar structure from
microlensing observations: analysis of time-dependent light curves
(e.g. \citealt{wwt00}; \citealt{k04}), or single epoch observations (\citealt{wms95};
\citealt{bfww08}). Both rely upon a lens model, which is used to
generate magnification maps that describe the fluctuation of
magnification induced by the microlenses, projected onto the source
plane. The lens model is defined in terms of the 
(external) shear, $\gamma$, and the convergence, $\kappa$.  Shear 
describes the distortion applied to images and is assumed to be dominated 
by matter in the host galaxy, however any matter along the line of sight 
to the source can contribute.  Convergence 
describes the focussing power of the (macro-)lens, and includes 
contributions from both smooth matter, $\kappa_s$, and 
compact objects, $\kappa_*$,  according to 
$\kappa = \kappa_s + \kappa_*$.  We define the smooth matter fraction to be
\begin{equation}
s = \kappa_s/\kappa.
\end{equation}

Once a set of model parameters have been defined, a magnification map 
can be generated using the gravitational lens equation in the following form:
\begin{equation}
{\mathbf y} = \left(
\begin{array}{cc}
1 - \gamma & 0 \\
0 & 1 + \gamma
\end{array}
\right) {\mathbf x} - \kappa_s {\mathbf x}
- \sum_{i=1}^{N_*} m_i \frac{({\mathbf x} - {\mathbf x}_i ) }
{\vert{\mathbf x} - {\mathbf x}_i\vert^2}
\label{eqn:sigma}
\end{equation}
which relates the two-dimensional location of a light 
ray, ${\mathbf x}$, with a source position, ${\mathbf y}$,
for
\begin{equation}
N_* = \frac{\kappa_* A}{\pi \langle M \rangle}
\label{eqn:Nstar}
\end{equation}
compact lenses distributed in an ellipse of area $A$, at lens 
plane coordinates, ${\mathbf x}_i$, and with mean lens mass, 
$\langle M \rangle$. 

Using Equation (\ref{eqn:sigma}) to shoot light rays at known
positions backwards through the lens plane to the source plane, 
we can count the number of rays reaching each pixel, $N_{ij}$. 
Comparing these values to the average number of light rays 
per pixel if there were no lensing, $N_{\rm rays}$, we obtain estimates 
for the per pixel magnification:
\begin{equation}
\mu_{ij} = N_{ij}/N_{\rm rays}.
\end{equation}

For large $N_*$ and large total number of light rays, the direct computation 
of Equation (\ref{eqn:sigma}) on a single-core CPU is not feasible.   The
hierachical tree-code (\citealt{wambsganss90}; \citealt{wambsganss99}) overcomes 
this limitation by replacing the $N_*$ lenses with a reduced number
of higher-mass pseudo-lenses.  A similar approach is used in the Fourier method \citep{k04},
where the lens and source planes are treated in a periodic fashion, which helps
reduce the total number of light rays required,  and also removes the
edge effects, making more of the magnification map usable for analysis.  The tessellation
approach (\citealt{Mediavilla06}; \citealt{Mediavilla11}) reduces the total number 
of light rays by using a sophisticated mapping
of polygonal regions from the lens plane to the source plane.  

While all of these algorithms
lead to improved computational efficiency, they do this at the expense of algorithmic (and
hence implementation) simplicity. The high level of parallelism in the
direct ray-shooting algorithm, coupled with the arithmetic intensity inherent
in Equation (\ref{eqn:sigma}), are ideal attributes for implementation on 
a GPU.  For details of the GPU-D code, including optimisation issues and timing tests, 
see \citet{thompson+10}.  For timing and accuracy comparisons between GPU, hierarchical 
tree codes and parallel large data codes, see \citet{bate+10}.  \citet{Mediavilla11} report on a timing comparison between an improved implementation 
of the tessellation method and the GPU-D results from \citet{bate+10}: they find a two order
of magnitude improvement in computational efficiency, but the total run-times of the two 
approaches were comparable.  

As the three (currently) CPU-only approaches for generating magnification maps
share a high-degree of parallelism, particularly through the need to calculate deflections
of many light rays, opportunities exist for their implementation on massively parallel
architectures.   Such work is beyond the scope of this present paper.  We adopt 
the viewpoint
that we have a working GPU ray-shooting code, and have access to a $\sim 100$ 
teraflop/s supercomputing
facility, gSTAR, and hence present what could be achieved now.   Any progress in reducing the total
computation time should be traded off with the additional amount of time required to implement
an alternative, more complex, solution \citep{fluke+11}.

\subsection{Map size and resolution}
\label{sec:determine}
In microlensing simulations, we run into the usual conflict between map size 
(measured in units of the Einstein radius) and map resolution. We need maps that 
have sufficient resolution to resolve the very inner regions of quasar accretion discs
($\sim10^{14}$cm; see for example the analysis of RXJ1131-1231 in \citealt{dai+10}), 
and sufficient size to map out quasar broad emission lines ($\sim10^{17}$cm for the
CIII]/CIV/MgII lines; see for example the microlensing analysis of \citealt{wow05}, 
or the reverberation mapping analysis of \citealt{kaspi+07}). One option is
to provide two sets of maps with sizes and resolutions tuned to the extreme physical 
scales for the accretion disk and BELR.  In light of recent multi-wavelength 
studies of accretion disk 
sizes and structures (\citealt{pmk08}; \citealt{morgan+10}; \citealt{blackburne+11};
\citealt{mosquera+11a}),  the approach we adopt is a single map that spans both physical
scales. This permits the use of self-consistent quasar models for studying
future, simultaneous multi-wavelength observations extending from accretion disk to 
BELR scales.

To set the pixel resolution required, we take the last stable orbit of
a Schwarzschild black hole of mass, $M_{bh}$:
\begin{equation}
R_{ISCO} = \frac{6GM_{bh}}{c^2} = 8.86 \times 10^{13} \left(\frac{M_{bh}}{10^8\rm{M_\odot}}\right)\rm{cm},
\end{equation}
where $G$ is the gravitational constant and $c$ is the speed of light.
We define this value for a $10^8\rm{M_\odot}$ black hole as the fiducial radius, 
$R_{ISCO}^{fid}$. For a fixed map size, the pixels required to
reach exactly this resolution will vary from system to system, 
as the Einstein radius depends on the angular diameter distances, $D_{ij}$, 
between the observer, $o$, lens, $d$, and source, $s$.  Projected onto the 
source plane, the Einstein radius is:
\begin{equation}
\eta_0 = \sqrt{\frac{D_{os} D_{ds}}{D_{od}}\frac{4G \langle M \rangle}{c^2}},
\label{eqn:er}
\end{equation}
for a microlens with mean mass, $\langle M \rangle$. For a background quasar at redshift $z=2.0$, a lens at $z=0.5$, and a mean microlens mass $\langle M \rangle = 0.3\rm{M_\odot}$, the Einstein Radius is $2.87\times10^{16}\rm{cm}$.

\begin{figure}
\centering
\includegraphics[angle=270, scale=0.5]{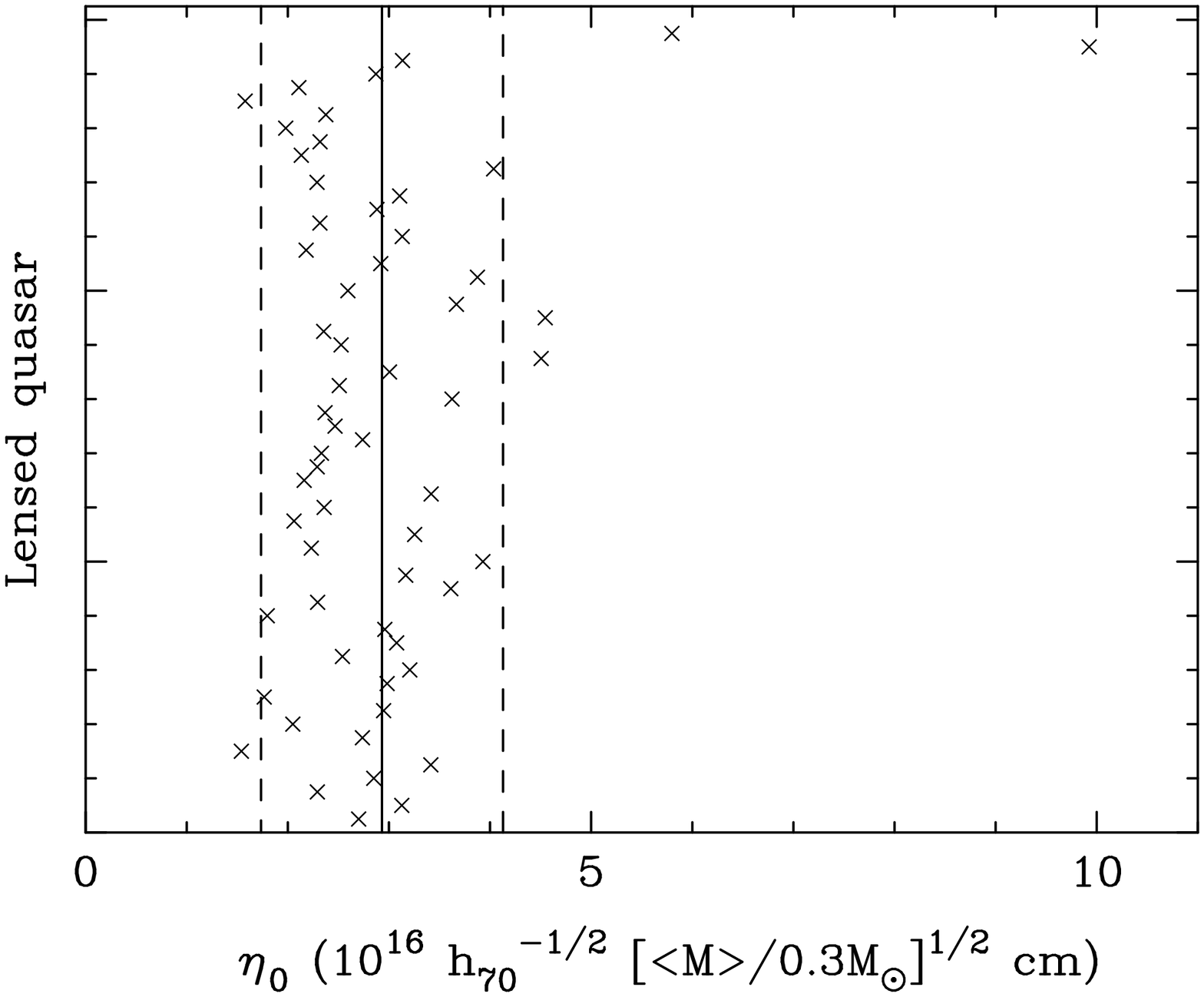}
\caption{The Einstein radius projected on to the source plane
  $\eta_0$ for each of the 59 lensing systems with both source and
  lens redshifts on the CASTLES Survey webpage
  (http://cfa-www.harvard.edu/glensdata/). The Einstein radius is
  calculated for a $0.3{\rm M}_\odot$ lens. The solid line indicates
  the mean $\eta_0$, and the dashed lines the standard deviation.\label{fig:mean_er}}
\end{figure}

How, then, do we determine the required number of pixels? 
We start by considering 59 of the currently known lensing systems with
redshifts available for both source and lens using data from the 
CASTLES Survey (this is a more conservative choice than \citealt{mosquera+11b}, who used estimates of lens redshifts to build up a sample consisting of 87 systems). Choosing $\langle M \rangle = 0.3 {\rm M_\odot}$, we 
calculate the mean Einstein radius $\langle \eta_0 \rangle$ and its standard deviation to be: 
\begin{equation}
\langle \eta_0 \rangle = (2.93 \pm 1.20)\times 10^{16} h_{70}^{-1/2} \left(\frac{\langle M \rangle}{0.3{\rm M}_\odot
}\right)^{1/2} \mbox{cm.}
\end{equation}
The Einstein radius of each system, along with the mean and standard deviation, are plotted in Figure \ref{fig:mean_er}. To achieve our target pixel resolution, $R_{ISCO}^{fid}$, our maps would require an average of $\sim 330$ pixels per $\eta_0$.

The two systems with unusually large Einstein radii are Q2237+0305 and MG1549+3047. In both cases, the large Einstein radii are a result of lensing galaxies at unusually low redshifts: $z_d = 0.04$ for Q2237+0305, and $z_d = 0.11$ for MG1549+3047. 
The latter is also an extended source seen as a radio ring \citep{lehar+93}, 
and is not known to exhibit microlensing, so it is an atypical object.
Removing these two systems essentially does not change the measured mean 
Einstein radius, but reduces the standard deviation by a factor of 1.79.
We make no comment here on selection effects that might bias this measurement; we simply note that 
the known sources are a reasonable place to start. Existing gravitationally lensed quasars have 
been discovered using a wide variety of techniques (see \citealt{kochanek06}), which are 
likely to be developed and refined in the era of large synoptic surveys.

Next, we determine the preferred physical size for the magnification maps. 
There are two main considerations here: that they be large enough to 
conduct simulations of quasar BELRs, and that they are wide enough to
allow long light curves to be simulated. 

Assuming the quasar H$\beta$ BELR radius-luminosity relation \citep{bentz+09}, $R_{BELR} \propto L^{0.519}$, holds for high-luminosity, high-redshift quasars (see \citealt{kaspi+07}), we can estimate the expected BELR radius. \citet{mosquera+11b} performed this calculation for their sample of 87 lensed quasars; for the 44 that overlap with our sample, the estimated mean H$\beta$ BELR and its standard deviation are $(1.71 \pm 1.47)\times 10^{17} \rm{cm}$, or $(5.84 \pm 5.01) \langle \eta_0 \rangle$.

Under the usual criterion that sources with sizes smaller than or approximately equal to the Einstein radius can be significantly microlensed, this suggests that H$\beta$ BELRs are generally too large to undergo significant microlensing. However, the BELR is known to be stratified, with higher ionisation lines emitted closer to the central source. In NGC 5548, for example, the CIV line is emitted from a region a factor of two smaller than the H$\beta$ line (see for example \citealt{peterson+91}; \citealt{dietrich+93}). Microlensing estimates of the size of the CIII] emission region in the lensed quasar Q2237+0305 suggest a radius similar to the Einstein radius for that source (\citealt{wow05}; \citealt{odowd+10}; \citealt{sluse+11}). In line with the microlensing analyses of \citet{abajas+02} and \citet{li04}, we therefore assume an approximate outer radius for the high ionisation BELR of $1\eta_0$.

A map with side length $50\eta_0$, convolved with a $1\eta_0$ radius
BELR, would provide $\sim500$ statistically independent
data points [taking into account edge effects due to convolution -- these can be avoided
using the Fourier method \citep{k04}]. This
drops dramatically if the BELR is much larger than $1\eta_0$, however
repeat generation of statistically independent maps allows us to
compensate for this drop off. A $50^2\eta_0$ map would require
$\sim16500^2$ pixels in order to resolve $R_{ISCO}^{fid}$.

Fortunately, it appears that the effective velocities of lensed
quasars are likely to be small enough that a map large enough to enable simulations of quasar BELRs will also be large enough to allow adequate sampling of microlensing
lightcurves. \citet{mosquera+11b} estimated that the average Einstein radius crossing time (the time taken for a source to travel a single Einstein radius) for a sample of 90 gravitationally lensed quasars peaks at $\sim23$ years, with a range of 8 to 44 years, assuming a mean microlens mass of $\langle M \rangle = 0.3{\rm M}_\odot$. We can therefore simulate observations of duration $\sim10$ years comfortably with a magnification map large enough for BELR simulations.

\subsection{Timing}

To further refine our choice of parameters, we investigate anticipated
run times. Using the empirical relationship, $T_{\rm GPU}$, 
in \citet{thompson+10} for a 4-GPU NVIDIA S1070 Tesla unit:
\begin{equation}
T_{\rm GPU} = 1.4 \times 10^4
\left(\frac{N_{\rm pix}}{4096^2}\right) \left(\frac{N_*}{10^6}\right)
\left(\frac{N_{\rm rays}}{100}\right)  {\rm sec}
\label{eqn:time}
\end{equation}
we can estimate the likely run time for any given magnification map.
We have not estimated times for the hierarchical code, but note that we 
expect them to be longer overall based on Figure 3 of \citet{bate+10}.

\citet{thompson+10} obtained a 
sustained processing performance of 1.28 teraflop/s on the S1070, which 
is approximately half the nominal peak performance (2.488 teraflop/s).
We note that quoted peaks are rarely achieved in scientific 
computations, as they typically require more than a dual-issued multiply
and addition per GPU clock cycle.  Assuming this half-peak performance
can also be obtained with gSTAR$_{100}$ (gSTAR$_{600}$), we can expect 
approximately 50 (300) teraflop/s performance utilising the entire 
facility -- an optimistic viewpoint that ignores real-world scheduling issues 
and the needs of other gSTAR users.
  
Following Bate et al. (2010), we set $N_{\rm rays}=1000$, as this was found 
to be sufficient for the parameters explored in that paper, and $N_{\rm pix}$ is 
allowed to vary in line with the discussion in Section \ref{sec:determine}.  
The remaining unknown in Equation (\ref{eqn:time}) is the number of 
microlenses, $N_*$, for each ($\kappa,\gamma$) pair, which we calculate 
from Equation (\ref{eqn:Nstar}). 

\begin{figure}
\epsscale{.60}
\plotone{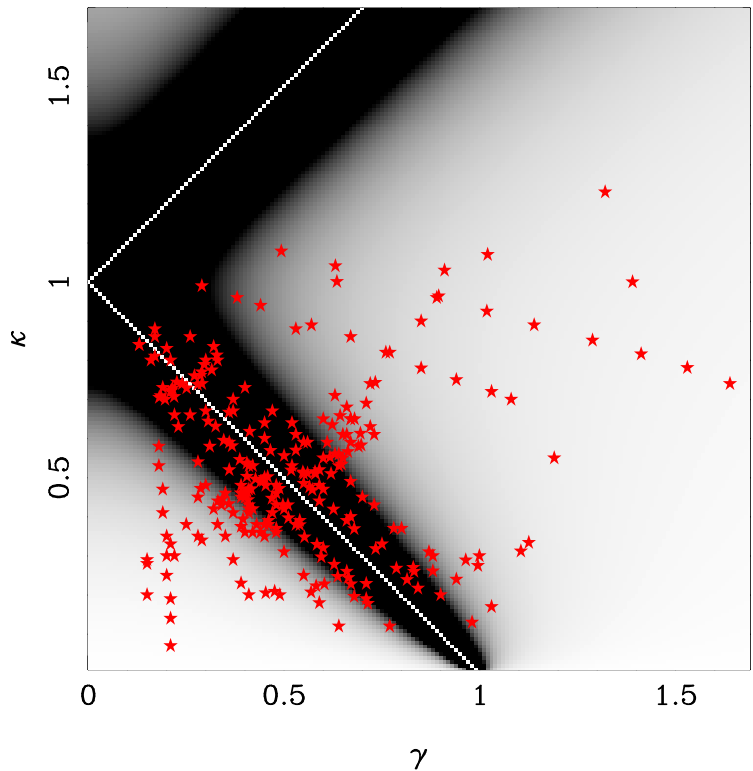}
  \caption{A contour plot showing simulation time for a magnification
    map with a given combination of $\kappa$ and $\gamma$. The colour
    gradient ranges from 0 hours (black) to 24 hours (white). Overlaid
    are the existing models taken from the literature (Table
    \ref{tab:models}). Times were calculated using Equation \ref{eqn:time},
    assuming maps with a side-length of $25\eta_0$, $N_{\rm pix} =
    10,000^2$ and $N_{\rm rays}=1000$.\label{fig:time_contour}}
\end{figure}

We also need to determine the total number of ($\kappa$, $\gamma$) combinations to 
cover parameter space. To avoid making any assumptions about the particular lens system each parameter
combination represents, we suggest that unifom tiling of $\kappa$--$\gamma$ parameter space is the most 
conceptually appealing approach.  It is likely that a single ($\kappa,\gamma$) combination will 
be relevant for more than one lensing system due to degeneracies in $\eta_0$.  
By examining published macromodel parameters for known microlensed quasars (see 
Table \ref{tab:models} and red symbols in Figure \ref{fig:time_contour}) we 
see that choosing a maximum $\kappa = \gamma = 1.7$ gives us good coverage.
We choose uniform increments $\Delta \kappa = \Delta \gamma = 0.01$, consistent
with the accuracy of most of the quoted macrolens models.  This results in
a $170 \times 171$ sampling of $\kappa$--$\gamma$ parameter space ($\kappa=0.0$ is excluded, as in this case there are no microlenses).

We obtain timing results for 10 values of the smooth matter fraction $s = \kappa_s/\kappa$ per ($\kappa$, $\gamma$) combination. The smooth matter fraction ranges from $0 \leq s < 1$ in increments of $\Delta s = 0.1$. For a given ($\kappa$, $\gamma$) combination, increasing the smooth matter fraction decreases the number of stars $N_*$. This is accounted for in our timing results.

We note that although we have attempted to collate all the ($\kappa$, $\gamma$) combinations in the literature for Table \ref{tab:models}, the list may not be complete. In addition, many authors publish enough information on lens macromodels to extract convergences and shears, but do not quote the values explicitly (see for example Table 6 in \citealt{blackburne+11}). We do not include such models in our analysis.

\begin{figure}
\epsscale{.60}
\plotone{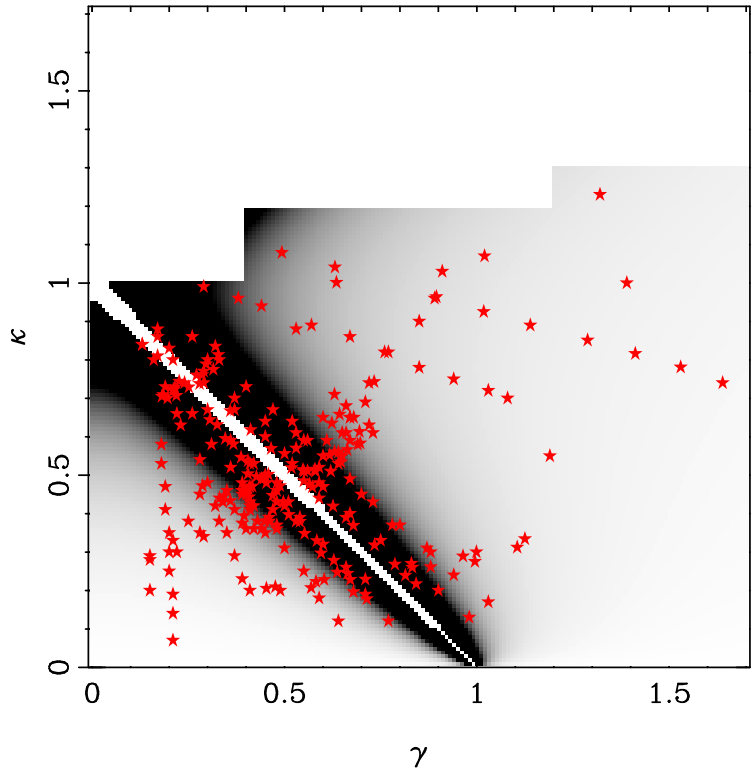}
  \caption{A contour plot showing simulation time for a magnification
    map with a given combination of $\kappa$ and $\gamma$, with maps
    that require simulation time greater than 1 gSTAR$_{100}$ day masked out. The colour
    gradient ranges from 0 hours (black) to 24 hours (white). Overlaid
    are the existing models taken from the literature (Table
    \ref{tab:models}). Times were calculated using Equation \ref{eqn:time},
    assuming maps with a side-length of $25\eta_0$, $N_{\rm pix} =
    10,000^2$ and $N_{\rm rays}=1000$.\label{fig:time_mask}}
\end{figure}

Timing contour plots are provided in Figures \ref{fig:time_contour}
and \ref{fig:time_mask}. The grayscale in both figures ranges from 0
hours (white) to greater than or equal to 24 hours (black). Note that
these are actual hours, not device hours.\footnote{These are based on our
estimate of processing on the entire cluster at 50 (300) teraflop/s, 
so that  1 gSTAR$_{100}$ (gSTAR$_{600}$) hour corresponds to $\sim10^{17}$ ($\sim10^{18}$)  computations in one real hour.}  Overplotted in both figures
(red stars) are the models from the literature (see Table
\ref{tab:models}). Figure \ref{fig:time_contour} shows the complete
timing; the white line indicating points where $(1-\kappa-\gamma)$ or $(1-\kappa+\gamma)$ equal zero corresponds 
to magnification maps that are formally infinite in size, and hence
are excluded from our computation. Figure \ref{fig:time_mask} 
shows timing coverage with regions requiring $>1$ gSTAR$_{100}$ day to process 
masked out.

In Table \ref{tab:options}, we present six possible strategies for
a GPU-enabled microlensing parameter survey. The first, labelled `Complete',
assumes that there are no limitations on the time we can run
the survey. Parameter space is completely covered, at the map size and
resolution that best suits the large microlensing simulations we would
ideally like to run (see Section \ref{sec:determine}). Obviously, this
parameter set is impractical; it would require running gSTAR$_{100}$ (gSTAR$_{600}$) 
non-stop for $\sim115$ ($\sim19$) years!

In the `Realistic' strategy, we sacrifice both map size and pixel
resolution in order to speed up the time to complete the
simulations. In addition, we exclude from this strategy any
magnification map that requires more than 1 gSTAR$_{100}$ (gSTAR$_{600}$) day to simulate. 
This amounts to 739 (215) excluded combinations of $\kappa$ and $\gamma$, 
or $\sim4$ ($\sim1$) percent of the total number of simulations. The simulation time is a 
more reasonable $\sim414$ gSTAR$_{100}$ days ($\sim167$ gSTAR$_{600}$ days).

Simulation strategy `Stage 1' gives an indication of the time required to
produce one magnification map for each of the model parameters in
Table \ref{tab:models}. All of the models in that table, representing
most (if not all) microlensing models in the literature, would be
simulated in $\sim122$ gSTAR$_{100}$ days ($\sim40$ gSTAR$_{600}$ days) using GPU-D. 
Starting with this strategy, successive processing stages can then fill in the rest 
of parameter space, moving outwards in regions centred around the 
current known models (Stages 2--4). 

We note that a significant amount of simulation time is spent in maps where $(1-\kappa-\gamma)$ or $(1-\kappa+\gamma)$ are close to zero. We have chosen not to include maps with simulation times greater than 1 gSTAR day in our timing. If this cutoff were reduced to 1 gSTAR hour, the `Realistic' strategy would take $\sim80$ gSTAR$_{100}$ days ($\sim34$ gSTAR$_{600}$ days), with 1617 (657) fewer ($\kappa$, $\gamma$) combinations.

\subsection{Data products}
Considering the total (unmasked) $\kappa$--$\gamma$ parameter space 
sampled $170 \times 171 = 29070$ times for a single smooth matter value, and 
with no repeat maps, the total uncompressed data volume is $\sim 53$ TB.   With 
consideration of appropriate data types, file formats, the use of 
compression, and an appropriate database infrastructure to handle 
queries and map retrieval, we assert that data storage and service is
reasonable. 

Magnification maps do not have negative values, so we can use 
an unsigned data type.  For the parameters we propose, an unsigned 
short integer (16-bit)
should be sufficient in most cases (for typical $N_{\rm rays} \sim 1000$), 
as this gives us a range from $0 \dots S_{\rm max} = 65536$ discrete 
values.  If there are cases where individual pixel values are 
$N_{ij} > S_{\rm max}$, 
it is unlikely that there are $S_{\rm max}$ distinct magnification values. 
In such a case, pixel values can be remapped into the range 
$0\dots S_{\rm max}$, and an index array can be built and stored as 
metadata.  In uncompressed form, and writing maps in a binary file 
format, a single $N_{\rm pix} = 10000^2$ map requires 191 MB of 
storage.\footnote{We use computing notation where 1 MB = 1 megabyte = 
1024 kilobytes, and similarly for gigabyte (GB), terabyte (TB) and 
petabyte (PB).}

To examine the likely effects of compression, we generate ten maps 
with randomly assigned values (up to $S_{\rm max}$) for each pixel.  
We then trial two standard Unix compression tools, {\tt gzip} and 
{\tt bzip2}, and record the compression ratio and the times to compress 
and decompress.  We find that {\tt gzip} compresses the random maps 
consistently to 171 MB with compression and decompression times
below 16 sec and 5 sec respectively.\footnote{Measured on a 2.8 GHz,
Intel Core 2 Duo, MacBook Pro, so these are upper limits.} Compare this 
with 157 MB file sizes for {\tt bzip2}, and $<38$ sec for 
compression and $<17$ sec for decompression.  

We expect compression of real maps to be somewhat better than random
pixels, as the range of per pixel values should be lower than a uniform
distribution over $0...S_{\rm max}$.  The presence of caustic structures means 
that there are sequences of repeated values in the maps, which tends to aid 
with the run-length encoding of {\tt gzip}, and the block-based 
compression of {\tt bzip2}.  We tested compression of a magnification
map with $\kappa = 0.416$, $\gamma = 0.471$, corresponding to the $\beta=1.0$ 
model for image $A$ in \objectname{Q2237+0305} in \citet{wms95}.   For
a $10000^2$-pixel map, and $N_{\rm rays} = 1000$, the count range
was  $15 \leq N_{ij} \leq 7060$.  
In this case, {\tt gzip} and {\tt bzip2} reduced the 
file size to 113 MB and 80 MB respectively. Times to compress and uncompress
the map were 23 sec and 5 sec for {\tt gzip} and 26 sec and 13 sec for 
{\tt bzip2}.  Despite the slightly longer decompression times, which is
actually the more critical factor for accessing data stored on disk 
in compressed
form, our testing suggests that {\tt bzip2} is the preferred
alternative, with likely compression ratios between $40$ per cent and $78$ per 
cent. These estimates are consistent with the compression ratios 
of 2.5 (40 per cent) to 3 (33 per cent) for OGLE data reported by \citet{pk10a}.

The overhead in decompressing any given map for analysis 
is small compared to the time taken to generate the map or to transfer 
large numbers of maps over Ethernet.   As the cost per TB of 
data storage continues 
to decrease at a rapid rate, we conclude that data storage and access, 
even with repeat maps and sampling of smooth matter values, is not 
a significant concern. An efficient, Web-based data delivery mechanism 
would be required
to serve sets of maps on the basis of user requests, and we are currently
considering options.

\section{First Application}
\label{sec:first_app}
We now present an example of one possible use for a microlensing parameter survey. 
In this section we analyse the effect of varying lensing parameters 
(specifically $\kappa$ and $\gamma$) on accretion disc constraints. 

As discussed in Section 1, lens macro-models are not uniquely determined. In systems where a wealth of observational data is available, complex models of the mass distribution in the lensing galaxy can be constructed (see for example the Q2237+0305 model of \citealt{trott+10}). More often, relatively simple parametric models are used to describe the lens mass distribution. These models are typically found to fit the lensed image positions adequately within observational errors, although additional mass components (perhaps due to other nearby galaxies, e.g. \citealt{pooley+06}) are sometimes required.

The main uncertainty in these models is the percentage of the surface density in smoothly distributed matter. This uncertainty has been included in many microlensing analyses, either as a free parameter in accretion disc measurements, or as the parameter of interest in analyses of lens dark matter fractions (see especially \citealt{mediavilla+09}; \citealt{bate+11}; \citealt{pooley+11}).

This uncertainty is typically confronted in one of two different ways. In the first technique, the galaxy is modelled with a de Vaucoulers profile embedded in a Navarro-Frenk-White (NFW) halo (e.g. \citealt{morgan+06}; \citealt{pmk08}; \citealt{chartas+09}; \citealt{morgan+10}; \citealt{dai+10}). The mass to light ratio is varied, typically from 1 (no NFW halo) to 0.1 (NFW halo dominates the mass distribution). This results in a sequence of lens macro-models, from which microlensing convergences and shears can be extracted.

In the second technique, a single macro-model is produced (usually a singular isothermal ellipse with external shear) to describe the lensing galaxy (e.g. \citealt{bfww08}; \citealt{mediavilla+09}; \citealt{blackburne+11}). A single set of microlensing convergences and shears are obtained from this model. The unknown smooth matter fraction is then accounted for by allowing the smooth $\kappa_s$ and stellar $\kappa_*$ components of the convergence to vary, subject to the constraint $\kappa = \kappa_s + \kappa_*$. We have assumed this second technique in our discussion on survey design, however we note that the magnification maps generated by such a survey are equally useful regardless of lens modelling technique.

Since a variety of lens modelling techniques are used, each providing acceptable fits to lensed image positions, it is reasonable to ask whether source quasar parameters can be uniquely determined from microlensing observations. We present a first attempt at a comparison between accretion disc constraints obtained using different lens models here.

We obtain our accretion disc constraints using a single-epoch imaging
technique very similar to that presented in \citet{bfww08} and
\citet{fbw09}. A similar analysis could (and should) be
undertaken for the lightcurve technique \citep{k04}, but is beyond the
scope of this paper.

As an additional
benefit, this analysis offers the first opportunity for a
direct comparison between accretion disc parameters obtained using the
lightcurve technique and the single-epoch imaging
technique. The observational data we use, consisting of pseudo-broadband
photometry of the  lensed quasar \objectname{Q2237+0305}, was presented and analysed
using the lightcurve technique in \citet{eigenbrod+08b}.

\subsection{Method}
The single-epoch microlensing simulation technique we use here was
first presented in \citet{bfww08}. The goal is to constrain both the
radius of the accretion disc $\sigma_0$ (at a particular wavelength $\lambda_0$), and the
power-law index $\zeta$ relating accretion disc radius $\sigma$ to observed
wavelength $\lambda$. In other words, we model the quasar accretion disc of
\object{Q2237+0305} using the following relationship:

\begin{equation}
\sigma = \sigma_0\left(\frac{\lambda}{\lambda_0}\right)^\zeta
\end{equation}
where $\lambda_0$ is taken to be the central wavelength of the bluest
filter in which observations were taken. 

The full simulation technique will
not be repeated here; it is described in full detail in Section 3 of
\citet{bfww08}. There are three key differences between the analysis
presented in that paper, and the analysis presented here. First, our
\object{Q2237+0305} magnification maps have a fixed microlens mass of $1{\rm M}_\odot$. Secondly, we use logarithmic Bayesian priors for source radius $\sigma$, rather
than the constant priors used in our previous analyses. Logarithmic priors ensure that the ratio of prior probability for two values of source size does not depend on the units chosen. This choice also allows us to compare more directly with previous work by other authors. And thirdly, we set the smooth 
matter percentage in the lensing galaxy at the image positions to zero.
This is reasonable because the lensed images lie in the bulge of the lensing galaxy, 
where stars are expected to dominate the mass distribution.

The result of this analysis technique is an a posteriori probability distribution for radius of the quasar accretion disc in the bluest filter $\sigma_0$, and power-law index $\zeta$ relating accretion disc radius to observed wavelength. For our purposes here, we will compare the extracted 68 per cent confidence limits on those two parameters.

We used lensing models for \object{Q2237+0305} from two sources: \citet{wms95} and \citet{k04}.
In \citet{wms95}, the authors modelled four lensing systems with a simple
family of lensing galaxy models, parameterised by the power-law index $\beta$
of the lensing potential with respect to radius $r$. Microlensing
parameters were quoted for three values of $\beta$: 0.5, 1.0 (the commonly used
singular isothermal sphere model) and 1.5. These three models allow us to test the effect of varying microlensing parameters on our accretion disc constraints. We also use the microlensing parameters from \citet{k04}, where the lensing galaxy was modelled as a singular isothermal ellipsoid (SIE) with external shear. This allows us to directly compare results obtained using our single-epoch imaging technique with results obtained using the lightcurve technique in \citet{eigenbrod+08b}. The parameters are found in Table \ref{tab:2237_models}. 

\subsection{Observational data}
The observational data were obtained
from \citet{eigenbrod+08b}, who conducted spectroscopic monitoring of
\object{Q2237+0305} using the VLT across a three year time period. The data, originally FORS1 spectra, were
deconvolved into a broad emission line component, a continuum emission
component, and an iron pseudo-continuum. The continuum emission was
fit with a power law of the form $f_\nu \propto
\nu^{\alpha_\nu}$. \citet{eigenbrod+08b} then split the
continuum into six wavelength bands, each with a width of 250\AA~in
the quasar rest frame, and integrated the continuum power-law in each
band. The result is pseudo-broadband photometry in six wavebands, with
contamination from broad emission lines and the iron continuum
removed. \citet{eigenbrod+08b} contains analysis of 39 of their 43
epochs using the lightcurve fitting technique of \citet{k04}. 

Here, we use two epochs of the \citet{eigenbrod+08b} data, separated by a year, and presented in Table
\ref{tab:2237_ratios}. We analyse one image combination: $B/A$. There is no reason in principle why the other images could not also be analysed. Here we choose $B/A$ both to cut down on simulation time, and to allow direct comparison with the \citet{eigenbrod+08b} results (which also only use the $B/A$ image combination). Probability distributions for $\sigma_0$ and $\zeta$ are
obtained for each epoch separately, and
then combined together to obtain our final results.

\subsection{Results and Discussion}
Results are quoted in units of the Einstein radius $\eta_0$ (see Equation \ref{eqn:er}). For \object{Q2237+0305}, this is $9.93
\times 10^{16}h_{70}^{-1/2}(\langle M
\rangle/0.3{\rm M}_{\odot})^{1/2}$cm. We present 68 per cent confidence limits on $\zeta$ and
$\sigma_0$ for all three models in Table \ref{tab:results}. The
\citet{eigenbrod+08b} results for their two cases (with and without a velocity prior) are provided for comparison.

The first thing to note is that within their errors, the constraints on both $\sigma_0$ and $\zeta$ agree for all models. This suggests that at the current level of accuracy, microlensing-based measurements of accretion disc parameters are likely to be consistent even when different lensing models are used.

However, there are some trends in our results which are worth discussing. If we look first at the three \citet{wms95} models, we see that as $\beta$ is increased from 0.5 to 1.5, so the peak value of $\zeta$ increases. It is difficult to disentangle what is causing this, however the most striking variation in the magnification maps as $\beta$ increases is an increase in the number of microlenses. The $\beta=1.5$ magnification maps are much more densely packed with caustic structure than their $\beta=0.5$ counterparts.

If we compare now the results obtained using the single-epoch imaging technique  (the \citealt{k04} lens model) to results obtained using the lightcurve technique (the \citealt{eigenbrod+08b} results), there are a few differences. Firstly, the lightcurve technique produces smaller errors. This isn't surprising; we analysed two epochs of data for the single-epoch imaging technique, whereas Eigenbrod and collaborators analysed 39 epochs of data. Secondly, for the same microlensing parameters the single-epoch technique results in slightly lower (although still consistent) values for $\zeta$ than the lightcurve technique. Whether this last effect is generically true would require more systems for comparison.

\section{Conclusion}
\label{sec:conc}
As a crucial step in preparing for this influx of new lensed systems that the synoptic
survey era promises to deliver, a large-scale exploration of microlensing convergence-shear 
parameter space is warranted. In this paper, we have described how a such a parameter space 
survey could be achieved using a modern GPU-based supercomputing facility, and our experiences
with the GPU-D code (\citealt{thompson+10}; \citealt{bate+10}).

Balancing simulation time and theoretical considerations, we suggest that the 
following is a suitable parameter set for such a survey:
\begin{itemize}
\item $0.01 \leq \kappa \leq 1.70$ with increment $\Delta \kappa = 0.01$
\item $0.00 \leq \gamma \leq 1.70$ with increment $\Delta \gamma = 0.01$
\item $0.0 \leq s < 1.0$ with increment $\Delta s = 0.1$
\item $N_{\rm rays} = 1000$
\item $N_{\rm pix} = 10000^2$
\item Map side length $=25\eta_0$
\end{itemize}
This results in a $170 \times 171$ sampling of $\kappa$--$\gamma$ parameter space for each $s$-value. The number of rays shot per pixel $N_{\rm rays}$ was found to be sufficient to produce accurate magnification maps using the GPU-D code in \citet{bate+10}. The pixel resolution and map size represent an acceptable compromise between the need to resolve the last stable orbit of a Schwarzschild supermassive black hole $R_{ISCO}^{fid}$, and the need for large enough maps to conduct microlensing simulations of quasar broad emission line regions and long light curves.

We collated microlensing models from the literature, and used them to suggest ways in which $\kappa$--$\gamma$ parameter space could be masked to reduce simulation time. In addition, a large portion of simulation time is spent in relatively few maps, where $(1-\kappa-\gamma)$ or $(1-\kappa+\gamma)$ are close to zero. Masking out any maps with projected simulation times greater than 1 day using the entire facility enables swifter completion of the survey, at the risk of missing some interesting areas of parameter space.  With these restrictions in place, we estimate a total simulation time of $\sim414$ days 
utilising the entirety of a 100 teraflop/s facility (gSTAR$_{100}$), or $\sim167$ days on a 600 teraflop/s 
facilty (gSTAR$_{600}$).  Realistic scheduling requirements for a high-demand supercomputer will 
increase the total wall-time required, so a staged strategy starting with parameters for the currently
known microlensed quasars, would support early science usage.

Finally, we have provided an example of one possible application for a large microlensing parameter survey. We used four different microlensing parameter sets to take a preliminary look at the systematic effects of varying lens models on accretion disc constraints in the lensed quasar Q2237+0305. Although all of the lens models produce results that are consistent within their errors, we note that there are suggestions that models with denser microlensing environments lead to shallower accretion disc temperature profile predictions.

\acknowledgments

This research was supported under Australian Research Council's Discovery 
Projects funding scheme (project number DP0665574).  We are grateful to 
Alex Thompson for his work on the GPU-D code.  Our referee provided very helpful
suggestions, that improved the quality of the work.  

{\it Facilities:} \facility{VLT:Kueyen (FORS1)}.

\clearpage

\clearpage

\clearpage

\clearpage

\clearpage

\begin{deluxetable}{lcc}
\tablewidth{0pt}
\tablecaption{Known microlens models and references.\label{tab:models}}
\tablehead{\colhead{System} & \colhead{$N_{\rm models}$} & \colhead{Ref.}}
\startdata
\object{HE 0047-1756} & 1 & (4) \\
\object{QJ 0158-4325} & 10 & (7) \\
\object{MG 0414+0534} & 3 & (13) \\
\object{HE 0435-1223} & 1 & (4) \\
\object{HE 0512-3329} & 1 & (4) \\
\object{SDSS 0806+2006} & 1 & (4) \\
\object{SDSS 0909+532} & 2 & (4), (5) \\
\object{SDSS J0924+0219} & 13 & (2), (4), (6) \\
\object{FBQ 0951+2635} & 1 & (4) \\
\object{QSO 0957+561} & 1 & (4) \\
\object{SDSS J1001+5027} & 1 & (4) \\
\object{SDSS J1004+4112} & 1 & (4) \\
\object{QSO 1017-207} & 1 & (4) \\
\object{HE 1104-1805} & 11 & (4), (7) \\
\object{PG 1115+080} & 5 & (4), (8), (13) \\
\object{RX J1131-1231} & 11 & (1), (4) \\
\object{SDSS J1206+4332} & 1 & (4) \\
\object{SDSS J1353+1138} & 1 & (4) \\
\object{H1413+117} & 4 & (4), (13) \\
\object{B J1422+231} & 1 & (4) \\
\object{SBS 1520+530} & 1 & (4) \\
\object{WFI J2033-4723} & 1 & (4) \\
\object{Q2237+0305} & 12 & (3), (9), (10), (11), (12), (13) \\
\enddata
\tablerefs{
(1) \citealt{dai+10}; (2) \citealt{keeton+06}; (3) \citealt{k04}; (4) \citealt{mediavilla+09}; (5) \citealt{mediavilla+11}; (6) \citealt{morgan+06}; (7) \citealt{morgan+08}; (8) \citealt{pooley+09}; (9) \citealt{rix+92}; (10) \citealt{swl98}; (11) \citealt{schneider+88}; (12) \citealt{wambsganss+94}; (13) \citealt{wms95}.}
\end{deluxetable}

\clearpage

\begin{deluxetable}{lccccl}
\tabletypesize{\scriptsize}
\rotate
\tablecaption{Simulation strategies.\label{tab:options}}
\tablewidth{0pt}
\tablehead{
\colhead{Strategy} & \colhead{Side length ($\eta_0$)} & \colhead{Side length (pixels)} & \colhead{Time
(gSTAR$_{100}$)} & \colhead{Time (gSTAR$_{600}$)} & \colhead{Notes}}
\startdata
Complete & 50 & 16500 & 115 years & 19 years & full parameter space simulation \\
Realistic & 25 & 10000 & 414 days & 167 days & \\
Stage 1 & 25 & 10000 & 122 days & 40 days & one map at each Table \ref{tab:models} model position \\
Stage 2 & 25 & 10000 & 177 days & 59 days & maps at Table \ref{tab:models} model positions $\pm0.01$ \\
Stage 3 & 25 & 10000 & 332 days & 140 days & maps at Table \ref{tab:models} model positions $\pm0.05$ \\
Stage 4 & 25 & 10000 & 392 days & 157 days & maps at Table \ref{tab:models} model positions $\pm0.1$ \\
\enddata
\tablecomments{Parameter space
  is tiled with a resolution of 0.01 in both $\kappa$ and
  $\gamma$, and and 0.1 in $s$. 
All times are estimated, and assume 100\% utilisation of gSTAR. 
The `Complete' strategy assumes no time constraints whatsoever, and
  is obviously impractical. All subsequent strategies exclude any
  parameter combinations that require greater than 1 device day per map to
  simulate, and mask out regions of parameter space containing no
  models (see Figure \ref{fig:time_mask}). Stages 1-4 represent
  progressive filling of parameter space towards the Realistic
  strategy. The listed simulation times are cumulative -- for example,
  Stage 2 takes 177 gSTAR$_{100}$ days to complete, but at the end of that
  time period Stage 1 will also have been simulated. }
\end{deluxetable}

\clearpage

\begin{deluxetable}{lccccl}
\tablecaption{Microlensing parameters for Q2237+0305. \label{tab:2237_models}}
\tablewidth{0pt}
\tablehead{
\colhead{Model} & \colhead{Image} & \colhead{$\kappa$} & \colhead{$\gamma$} & \colhead{Magnification $\mu$} & \colhead{Ref.}}
\startdata
$\beta = 0.5$ & $A$ & 0.228 & 0.603 & 4.3 & (1) \\
 & $B$ & 0.223 & 0.582 & 3.8 & \\
 & $C$ & 0.289 & 0.964 & -2.4 & \\
 & $D$ & 0.261 & 0.880 & -4.4 &  \\
$\beta=1.0$ & $A$  & 0.471 & 0.416 & 9.4 & (1) \\
 & $B$ & 0.464 & 0.406 & 8.1 & \\
 & $C$ & 0.552 & 0.617 & -5.6 & \\
 & $D$ & 0.515 & 0.579 & -10.0 &  \\
$\beta=1.5$ & $A$  & 0.728 & 0.215 & 36.5 & (1) \\
 & $B$ & 0.723 & 0.212 & 31.3 & \\
 & $C$ & 0.789 & 0.296 & -23.3 & \\
 & $D$ & 0.761 & 0.286 & -40.7 &  \\
SIE$+\gamma$ & $A$  & 0.394 & 0.395 & 4.7 & (2) \\
 & $B$ & 0.375 & 0.390 & 4.2  & \\
 & $C$ & 0.743 & 0.733 & -2.1 & \\
 & $D$ & 0.635 & 0.623 & -3.9 &  \\
\enddata
\tablerefs{(1) \citealt{wms95}; (2) \citealt{k04}.}
\end{deluxetable}

\clearpage

\begin{deluxetable}{lrrlll}
\tablecaption{Two epochs of observational $B/A$ and $C/A$ flux ratios
  for Q2237+0305. \label{tab:2237_ratios}}
\tablewidth{0pt}
\tablehead{
\colhead{Band} & \colhead{Emitted $\lambda_c$ (\AA)} & \colhead{Observed $\lambda_c$} & \colhead{$B/A$} &
\colhead{$C/A$} & \colhead{Date}}
\startdata
1 & $1625\pm125$ & $4379\pm337$ & $0.52\pm0.02$ & $0.31\pm0.03$ & 2005 November 11 \\
2 & $1875\pm125$ & $5053\pm337$ & $0.51\pm0.02$ & $0.33\pm0.03$ & 2005 November 11 \\
3 & $2125\pm125$ & $5727\pm337$ & $0.50\pm0.01$ & $0.34\pm0.03$ & 2005 November 11 \\
4 & $2375\pm125$ & $6401\pm337$ & $0.49\pm0.01$ & $0.35\pm0.03$ & 2005 November 11 \\
5 & $2625\pm125$ & $7074\pm337$ & $0.48\pm0.01$ & $0.37\pm0.03$ & 2005 November 11 \\
6 & $2875\pm125$ & $7748\pm337$ & $0.47\pm0.01$ & $0.37\pm0.02$ & 2005 November 11 \\
\tableline
1 & $1625\pm125$ & $4379\pm337$ & $0.33\pm0.02$ & $0.28\pm0.02$ & 2006 November 10 \\
2 & $1875\pm125$ & $5053\pm337$ & $0.34\pm0.02$ & $0.30\pm0.02$ & 2006 November 10 \\
3 & $2125\pm125$ & $5725\pm337$ & $0.35\pm0.01$ & $0.31\pm0.02$ & 2006 November 10 \\
4 & $2375\pm125$ & $6401\pm337$ & $0.36\pm0.02$ & $0.33\pm0.02$ & 2006 November 10 \\
5 & $2625\pm125$ & $7074\pm337$ & $0.37\pm0.02$ & $0.34\pm0.02$ & 2006 November 10 \\
6 & $2875\pm125$ & $7748\pm337$ & $0.37\pm0.01$ & $0.35\pm0.02$ & 2006 November 10 \\
\enddata
\tablecomments{Data obtained from \citet{eigenbrod+08b}. Photometry was extracted from spectra obtained with the FORS1 spectrograph on the Very Large Telescope (VLT) at the European Southern Observatory (ESO). Filter wavelengths are in the rest frame of the source quasar, located at a redshift of $z_s=1.695$. Following the \citet{eigenbrod+08b} numbering, the 2005 November 11 dataset is epoch number 17, and the 2006 November 10 dataset is epoch number 28.} 
\end{deluxetable}

\clearpage

\begin{deluxetable}{lcc}
\tablecaption{Accretion disc constraints for three Q2237+0305
  models.\label{tab:results}}
\tablewidth{0pt}
\tablehead{
\colhead{Model} & \colhead{Radius $\sigma_0$} & \colhead{Power-law index $\zeta$}}
\startdata
$\beta=0.5$ & $<0.15\eta_0$ & $0.79^{+0.63}_{-0.37}$ \\
$\beta=1.0$ & $<0.15\eta_0$ & $0.88^{+0.56}_{-0.37}$ \\
$\beta=1.5$ & $<0.16\eta_0$ & $1.40^{+0.83}_{-0.58}$ \\
\citet{k04} model & $<0.21\eta_0$ & $0.77^{+0.57}_{-0.34}$ \\
\citet{eigenbrod+08b} velocity prior & $0.16^{+0.12}_{-0.10}\eta_0$ & $1.2\pm0.3$\\
\citet{eigenbrod+08b} no velocity prior & $0.69^{+1.30}_{-0.60}\eta_0$ & $1.1\pm0.3$ \\
\enddata
\tablecomments{Constraints on radius of the accretion disc in
  \objectname{Q2237+0305} at $\lambda_{obs}=4379$\AA, and power-law
  index relating accretion disc radius and observed wavelength of
  emission. Constraints were obtained using the $B/A$ image
  combination. Results from \citet{eigenbrod+08b}, which were obtained
  using the lightcurve technique of \citet{k04}, are provided for comparison.}
\end{deluxetable}

\end{document}